\documentclass[12pt]{article}
\usepackage{amsfonts,latexsym}
\begin{document}
\newcommand{\e}{\epsilon} \newcommand{\ot}{\otimes}
\newcommand{\be}{\begin{equation}} \newcommand{\ee}{\end{equation}}
\newcommand{\ba}{\begin{eqnarray}} \newcommand{\ea}{\end{eqnarray}}
\newcommand{\tmod}{{\cal T}}\newcommand{\amod}{{\cal A}}
\newcommand{\bemod}{{\cal B}}\newcommand{\cmod}{{\cal C}}
\newcommand{\dmod}{{\cal D}}\newcommand{\hmod}{{\cal H}}
\newcommand{\s}{\scriptstyle}\newcommand{\tr}{{\rm tr}}
\newcommand{\einsop}{{\bf 1}}
\def\oR{R^*} \def\upa{\uparrow}
\def\R{\overline{R}} \def\doa{\downarrow}
\def\nn{\nonumber} \def\dag{\dagger}
\def\be{\begin{equation}}
\def\ee{\end{equation}} 
\def\bea{\begin{eqnarray}} 
\def\eea{\end{eqnarray}} 
\def\R{\check{R}} 
\def\ve{\epsilon}
\def\s{\sigma}
\def\th{\theta} 
\def\a{\alpha} 
\def\b{\beta}
\def\g{\gamma}
\def\h{\overline{h}}
\def\d{\delta}
\def\m{\eta} 
\def\l{\left}
\def\r{\right} 
\def\n{\tau}
\def\ua{\uparrow}
\def\da{\downarrow}
\newcommand{\case}[2]{\textstyle{#1\over#2}\displaystyle}
\newcommand{\reff}[1]{eq.~(\ref{#1})}
\centerline{\bf{\huge Exactly solvable quantum spin ladders  }} 
\centerline{\bf{\huge associated with the orthogonal  }}
\centerline{\bf{\huge and symplectic Lie algebras}} 
~~~\\
\begin{center}
{\large M.T. Batchelor$\,^a$, J. de Gier$\,^a$, J. Links$\,^b$ and 
M. Maslen$\,^a$}
\vspace{0.5cm}~~\\
$^a${\em Department of Mathematics, School of Mathematical Sciences,\\
The Australian National University, Canberra ACT 0200, Australia}
\vspace{0.5cm}~~\\
$^b${\em Centre for Mathematical Physics, Department of Mathematics,\\
The University of Queensland,  QLD 4072,  Australia \\
}
~~\\
~~\\
\end{center} 
\begin{abstract}
We extend the results of spin ladder models associated with the Lie 
algebras $su(2^n)$ to the case of the orthogonal and symplectic algebras
$o(2^n),\ sp(2^n)$ where $n$ is the number of legs for the system.  
Two classes  of models are found whose symmetry, either orthogonal or 
symplectic, has an explicit $n$ dependence. Integrability of these models
is shown for an arbitrary coupling of $XX$ type rung interactions and 
applied magnetic field term.
\end{abstract}
\vspace{1cm}
\begin{flushleft}  
\end{flushleft}

\vfil\eject
~~\\
The study of quantum spin ladders has recently been the focus of both
theoretical and experimental investigations \cite{dr}.
{}From the theoretical point of view, many researchers are now applying the 
procedures of the Quantum Inverse Scattering Method (QISM) and associated 
Bethe Ansatz techniques in order to obtain non-perturbative results for 
models of spin ladder systems. Originally this was undertaken in
\cite{fr1,fr2} by using the co-algebra structure of the Yangian algebra
(in fact a Hopf algebra) which underlies the applicability of the QISM.
In essence the co-algebra mapping allows the one-dimensional model to be
mapped to multiple copies by way of a homomorphism. As such, the
algebraic properties are preserved and consequently integrability is
maintained, allowing for analyses of these models by standard methods.

Subsequently, an alternative view point has been adopted in which the
extension from the one-dimensional system to the ladder model is
accomodated by an extension of the symmetry of the system. 
By such an approach, the ladder is considered 
as a quasi one-dimensional system whereby the rung interactions of the 
ladder are represented as local symmetry operators. In this manner, the rung 
interactions may be coupled to the bulk ladder model without violating 
integrability. This in turn again facilitates the application of  
many known techniques that have been developed in the rich field of integrable 
models in order to determine ground state properties, elementary 
excitation spectra and phase diagrams.

One of the first works in this latter direction is that of Wang \cite{w} 
in which 
an integrable formulation of a two-leg spin 1/2 model was given in terms of 
the maximal local (i.e. rung space) symmetry algebra of $su(4)$.  
In this model rung interactions are  coupled with arbitrary strength
and a phase diagram was obtained in terms of the coupling parameter. In
the same paper another model based on $su(3|1)$ supersymmetry was also
presented and it became  apparent that such an approach could be
 extended to incorporate other forms of symmetry.   
In \cite{bm1} it was demonstrated that the natural extension to $n$-leg
models gave rise to $su(2^n)$ symmetry. Other models have also been
considered using  a ``non-standard''
solution of the Yang-Baxter relation \cite{afw,bm2} and others  
based on $su(4)$ and $su(2|2)$ symmetry \cite{f} in the context of
fermion ladder systems.  
Extension of this approach to a  
$t-J$ ladder model which is also of significant physical interest can be
found in \cite{fk}.

In the spirit of \cite{bm1} our goal here is to identify $n$-leg ladder models
arising through the orthogonal and symplectic Lie algebras. We find two
distinct classes which are integrable for arbitrary coupling of $XX$
rung interactions and magnetic field terms. A curious feature of the
models we obtain is that the symmetry algebra has an explicit dependence
on the number of legs of the system.

Throughout, we will express all operators in terms of the Pauli matrices
\be
\s^x=\pmatrix{0&1 \cr 1&0},~~\s^y=\pmatrix{0&-i\cr i&0},
~~\s^z=\pmatrix{1&0\cr 0&1}
\ee
which act on a two-dimensional space that we denote $V$. Furthermore, 
we use a superscript $(l)$ to denote an operator acting on
the leg labelled by $l$ and use subscripts to label the rungs.

In order to construct a model on an $n$-leg ladder we begin by introducing
a metric for the $2^n$-dimensional space $W=V^{\ot n}$ by
\bea 
\a&=&\s^x\ot \s^y\ot \s^x\ot \s^y\ot ....\ot X \nn \\
&=&\prod_{l\,{\rm odd}}\l(\s^x\r)^{(l)} 
\prod_{l'\,{\rm even}}\l(\s^y\r)^{(l')} 
\eea
where $X=\s^x$ for an odd number of legs and $\s^y$ for the even case. 
The metric has the useful properties,
\be
\a=\a^{-1}=\a^{\dagger}=(-1)^{[n/2]}\a^t,
\ee
with $[n/2]$ being the integer part of $n/2$. For any basis 
$\{v_i\}_{i=1}^{2^n}$ of $W$, the set of matrices
\be A^i_j=e^i_j-\a e^j_i\a\label{algebra}, \ee
with the elementary matrices $e^i_j$ satisfying $e^i_jv^k=\delta^k_j
v^i$, in fact close to form a Lie algebra. The commutation relations
amongst the generators read
\be
[A^i_j,\,A^k_l]=\d^k_jA^i_l-\d^i_lA^k_j+\a^i_k\a^l_mA^m_j-\a^l_j\a^m_k
A_m^i
\ee
with implied summation over the repeated index $m$.
For the instance where $[n/2]$ is even these matrices realize the Lie 
algebra $o(2^n)$ while in the odd case it is $sp(2^n)$.

As in the $gl(2^n)$ invariant models we begin with the observation that the
 permutation operator on $W_i\ot W_j$ 
is expressible in terms of the Pauli matrices by 
\be
P_{ij}=\frac{1}{2^n}\prod_{l=1}^n\l(I+{\s}_i^{(l)}\cdot
\mathbf{\s}^{(l)}_j\r).
\ee
We now introduce a Temperley-Lieb operator \cite{mr1},
\be
Q_{ij}=(I\ot \a)P^{t_j}(I\ot \a^t),
\ee
with the properties, 
\bea Q^2&=&(-1)^{[n/2]}2^nQ \nn \\
Q_{i(i+1)}&=&Q_{i(i+ 1)}Q_{(i+1)(i+2)}Q_{i(i+1)} \nn \\
Q_{i(i+1)}&=&Q_{i(i+ 1)}Q_{(i-1)i}Q_{i(i+1)} \eea  
This operator on $W_i\ot W_j$ reads, when expressed in Pauli matrices,
\be
Q_{ij}=\frac{(-1)^{[n/2]}}{2^n}\prod_{l=1}^n\l(I-\s^{(l)}_i \cdot
M^{(l)}\s^{(l)}_j\r),
\ee
where the matrix $M^{(l)}$ is  diagonal  
with entries
\be
M^{(l)}={\rm diag} \l((-1)^l,\,(-1)^l, 1\r). 
\ee

It is well known that given any Temperley-Lieb operator there is an
associated integrable model. The operator defined by 
\be
\R(u)=I+\frac{(-1)^{[n/2]}\sinh u}{\sinh(\gamma-u)}Q,   
\ee
gives a solution of the Yang-Baxter equation,
\be 
\R_{12}(u)\R_{23}(u+v)\R_{12}(v)=\R_{23}(v)\R_{12}(u+v)\R_{23}(u),
\label{yb}
\ee
where the parameter $\gamma$ is determined by 
\be
\cosh\gamma=2^{(n-1)}.
\ee

Furthermore,  the set of operators $\{G_i=(-1)^{[n/2]}P_{i(i+1)},\,
E_i=Q_{i(i+1)}\}_{i=1}^{L-1}$  together give a representation of the
Birman-Wenzl-Murakami algebra \cite{bmw}.
Using the results of \cite{cgx}, we can now obtain the operator
\be
\R(u)=I+uP-\frac{u}{u+2^{(n-1)}-(-1)^{[n/2]}}Q,
\ee
which also gives a solution of the Yang-Baxter equation (\ref{yb}). 

By the standard approach, integrable models with periodic boundary 
conditions of the form 
\be 
H=\sum_{i=1}^{L-1}H_{i(i+1)} +H_{L1}, \label{ham} 
\ee  
can be obtained from the above solutions of the Yang-Baxter equation,  
where the local Hamiltonians $H_{ij}$ read
\be
H_{ij}=\l.\frac{d}{du} \R_{ij}(u)\r|_{u=0}.
\ee
For the Temperley-Lieb models we obtain 
\bea 
H_{ij}&=&\frac{(-1)^{[n/2]}}{\sinh\gamma}Q_{ij} \nn \\
&=&\frac{1}{2^n\sinh\gamma}\prod_{l=1}^n\l(I-\s^{(l)}_i\cdot
M^{(l)}\s^{(l)}_j\r), \label{ham1} 
\eea 
while in the Birman-Wenzl-Murakami case we have 
\bea 
H_{ij} &=&P_{ij}-\frac{1}{2^{(n-1)}-(-1)^{[n/2]}}Q_{ij} \nn \\
&=&\frac{1}{2^n}\prod_{l=1}^n\l(I+\s_i^{(l)}.\s_j^{(l)}\r)
 -\frac{(-1)^{[n/2]}}{2^n(2^{(n-1)}-(-1)^{[n/2]})}\prod_{l=1}^n
\l(I-\s^{(l)}_i\cdot M^{(l)}\s_j^{(l)}\r). \label{ham2} 
\eea 

By our construction, the above Hamiltonians  are  invariant with respect to
the algebra elements defined by (\ref{algebra}). 
Additionally, we can couple to these models $XX$ rung 
interactions and applied magnetic field terms 
\be
\case{1}{2}J\sum_{i=1}^L\sum_{l=1}^{n-1} h^{(l)}_{i}+\case{1}{2} B
\sum_{i=1}^L\sum_{l=1}^{n} g^{(l)}_{i},
\ee
where 
\bea h_i^{(l)}&=&\l(\s^x\r)_i^{(l)}\l(\s^x\r)_i^{(l+1)}+
\l(\s^y\r)_i^{(l)}\l(\s^y\r)_i^{(l+1)},\nn \\
g_i^{(l)}&=&\l(\s^z\r)_i^{(l)}.
\eea
For an even number of legs we can also impose periodic boundary conditions
on the rungs giving rise to a tube model as done in \cite{bm1}.
The rung interactions and magnetic field terms above commute with the 
Hamiltonians (\ref{ham1}, \ref{ham2}) as 
a consequence of the fact that they can be 
expressed in terms of the symmetry algebra elements (\ref{algebra}).

It is important to mention that the Bethe Ansatz solutions of the 
models discussed above are all known. Using the Temperley-Lieb
equivalence, the spectra of the above Temperley-Lieb models (in the
absence of rung interactions and field terms) coincide
with that of the $XXZ$ chain with the choice 
\be
\Delta= -2^{(n-1)},
\ee
where $\Delta$ is the $XXZ$ anisotropy. In this limit the
Temperley-Lieb ladder models are thus massive for all $n$.
For the Birman-Wenzl-Murakami class  the solutions  were originally
obtained by Reshetikhin using the analytic Bethe Ansatz approach
\cite{r}. More recently, these results have been rederived in an
algebraic fashion by Martins and Ramos \cite{mr2}. For the simplest
case, that of the two-leg model corresponding to $sp(4)$ symmetry, the
energy levels of the Hamiltonian are given by 
\be
E=L+\sum_{i=1}^{M_1}\frac{1}{u_i^2-1/4}+J(M_1-L)+B(M_1-2M_2),
\label{energy1}
\ee
where the parameters $u_i$ are solutions of the Bethe Ansatz equations
\bea -\l(\frac{u_i+1/2}{u_i-1/2}\r)^N&=&\prod_{j=1}^{M_1}\frac{
u_i-u_j+1}{u_i-u_j-1}.\prod_{k=1}^{M_2}\frac{u_i-v_k-1}{u_i-v_k+1} 
,~~~~~j=1,2,....,M_1\nn \\
-1&=&\prod_{j=1}^{M_1}\frac{u_j-v_i-1}{u_j-v_i+1}\prod_{k=1}^{M_2}
\frac{v_k-v_i+2}{v_k-v_i-2},~~~~~i=1,2,....,M_2. 
\eea
Above the parameters $M_1,\,M_2$ are restricted to the intervals
\bea 
&&0\leq M_1\leq2L, \nn \\
&&0\leq M_2\leq{\rm min}(M_1,L). \nn \eea 
The expression for the energy (\ref{energy1}) was derived using the
the following rung basis states,
in terms of which the rung interactions and field terms are diagonal, 
\bea
|0\rangle &=& \frac{1}{\sqrt{2}} \left( |\ua\da\rangle -
|\da\ua\rangle \right), \nn\\
|+\rangle &=& |\ua\ua\rangle, \nn\\
|-\rangle &=& |\da\da\rangle, \nn\\
|\pm\rangle &=& \frac{1}{\sqrt{2}} \left( |\ua\da\rangle +
|\da\ua\rangle \right).\nn \eea
These states may be interpreted as an empty site, occupation by a $+$ or a
$-$ particle and a bound state, respectively. An advantage of this 
identifcation is that we may now write
\bea M_1&=&\sum_{i=1}^Ln^+_i+n^-_i, \nn \\
 M_2&=&\sum^L_{i=1}n_i^- . \nn \eea
Note that 
the transformation $|0\rangle \leftrightarrow |\pm\rangle$, $|+\rangle
\leftrightarrow |-\rangle$, $B\rightarrow -B$, $J\rightarrow-J$, is a
symmetry of the system. Furthermore, the transformation 
$|0\rangle \leftrightarrow |-\rangle$, $|+\rangle
\leftrightarrow |\pm\rangle$, $B\leftrightarrow J$
also leaves the model invariant.

It remains to investigate the thermodynamic properties of the
ladder models. Without a magnetic field, it is easy to see that if $J$
is large and positive, $M_1=0$ for the ground state. In this case the
system is completely empty and any excitation is massive. Since the model
is critical at $J=0$, there is a phase transition at some critical
value $J_{\rm c}$, similar to that in the $su(4)$ case \cite{w}. By a
similar analysis it follows that for large but negative $J$ the system
is massive again, but now completely filled. At negative $J$ there
therefore exists another critical point, again similar to that of
\cite{w}. In principle one should be able to calculate the location of
these phase transitions exactly. However, little is known about the
Bethe Ansatz solutions for the quantum spin chains associated with the
orthogonal and symplectic Lie algebras.  

\begin{flushleft}
{\bf Acknowledgements}
\end{flushleft}
This work is supported by the Australian Research Council.

\end{document}